\documentclass[aps,showpacs,twocolumn,superscriptaddress,nofootinbib] {revtex4-1}  
\usepackage{graphicx,color}  
\usepackage{dcolumn}   
\usepackage{bm}        
\usepackage{amssymb} 
\usepackage{amsmath}
\usepackage{braket}
 \newcommand{\beq}{\begin{equation}}
\newcommand{\eeq}{\end{equation}}
\newcommand{\beqa}{\begin{eqnarray}}
\newcommand{\eeqa}{\end{eqnarray}}
\newcommand{\be}{\begin{equation}}
\newcommand{\ee}{\end{equation}}
\newcommand{\bea}{\begin{eqnarray}}
\newcommand{\eea}{\end{eqnarray}}
\newcommand{\sm}{\mathcal{S}}

\newcommand{\bnabla}{\bar\nabla}
\newcommand{\p}{\pi}
\newcommand{\sdfrac}[2]{\mbox{\small$\displaystyle\frac{#1}{#2}$}}
\begin{document}
\def\eg{{\it e.g.}}
\newcommand{\nc}{\newcommand}
\nc{\rnc}{\renewcommand}
\rnc{\d}{\mathrm{d}}
\nc{\D}{\partial}
\nc{\K}{\kappa}
\nc{\bK}{\bar{\K}}
\nc{\bN}{\bar{N}}
\nc{\bq}{\bar{q}}
\nc{\vbq}{\vec{\bar{q}}}
\nc{\g}{\gamma}
\nc{\lrarrow}{\leftrightarrow}
\nc{\rg}{\sqrt{g}}
\rnc{\[}{\begin{equation}}
\rnc{\]}{\end{equation}}
\nc{\nn}{\nonumber}
\rnc{\(}{\left(}
\rnc{\)}{\right)}
\nc{\q}{\vec{q}}
\nc{\x}{\vec{x}}
\rnc{\a}{\hat{a}}
\nc{\ep}{\epsilon}
\nc{\tto}{\rightarrow}
\rnc{\inf}{\infty}
\rnc{\Re}{\mathrm{Re}}
\rnc{\Im}{\mathrm{Im}}
\nc{\z}{\zeta}
\nc{\mA}{\mathcal{A}}
\nc{\mB}{\mathcal{B}}
\nc{\mC}{\mathcal{C}}
\nc{\mD}{\mathcal{D}}
\nc{\mN}{\mathcal{N}}
\rnc{\H}{\mathcal{H}}
\rnc{\L}{\mathcal{L}}
\nc{\<}{\langle}
\rnc{\>}{\rangle}
\nc{\fnl}{f_{NL}}
\nc{\gnl}{g_{NL}}
\nc{\fnleq}{f_{NL}^{equil.}}
\nc{\fnlloc}{f_{NL}^{local}}
\nc{\vphi}{\varphi}
\nc{\Lie}{\pounds}
\nc{\half}{\frac{1}{2}}
\nc{\bOmega}{\bar{\Omega}}
\nc{\bLambda}{\bar{\Lambda}}
\nc{\dN}{\delta N}
\nc{\gYM}{g_{\mathrm{YM}}}
\nc{\geff}{g_{\mathrm{eff}}}
\nc{\tr}{\mathrm{tr}}
\nc{\oa}{\stackrel{\leftrightarrow}}
\nc{\IR}{{\rm IR}}
\nc{\UV}{{\rm UV}}

\title{\Large \bf Broken Scale Invariance and the Regularization\\
of a Conformal Sector in Gravity with Wess-Zumino actions\\ }
\vspace{1.5cm}
 \vspace{0.3cm}
\vspace{1cm}
\author{Claudio Corian\`o }
\affiliation{\it Dipartimento di Matematica e Fisica, Universit\`{a} del Salento \\
and INFN Sezione di Lecce, Via Arnesano 73100 Lecce, Italy\\
National Center for HPC, Big Data and Quantum Computing}
\author{Mario Cret\`i  }
\affiliation{\it Dipartimento di Matematica e Fisica, Universit\`{a} del Salento \\
and INFN Sezione di Lecce, Via Arnesano 73100 Lecce, Italy\\
National Center for HPC, Big Data and Quantum Computing}
\affiliation{\it Center for Biomolecular Nanotechnologies,\\ Istituto Italiano di Tecnologia, Via Barsanti 14,
73010 Arnesano, Lecce, Italy\\}
\author{Matteo Maria Maglio}
\affiliation{\it  Institute for Theoretical Physics (ITP), University of Heidelberg\\
	Philosophenweg 16, 69120 Heidelberg, Germany}

\begin{abstract}
We elaborate on anomaly induced actions of the Wess-Zumino (WZ) form and their relation to the renormalized effective action, which is defined by an ordinary path integral over a conformal sector, in an external gravitational background. In anomaly 
-induced actions, the issue of scale breaking is usually not addressed, since these actions are obtained only by solving the trace anomaly constraint and are determined by scale invariant functionals. 
We investigate the changes induced in the structure of such actions once identified 
in dimensional renormalization (DR) when the $\epsilon=d- 4\to 0$ limit is accompanied by the dimensional reduction (DRed) of the field dependencies. 
We show that operatorial nonlocal modifications $(\sim\Box^\epsilon)$ of the counterterms are unnecessary to justify a scale anomaly. In this case, only the ordinary finite subtractions play a critical role in the determination of the scale breaking. This is illustrated for the WZ form of the effective action and its  WZ consistency condition, as seen from a renormalization procedure. Logarithmic corrections from finite subtractions are also illustrated in a pure $d=4$ (cutoff) scheme. The interplay between two renormalization schemes, one based on dimensional regularization (DR) and the second on a cutoff in $d=4$, illustrates the ambiguities of DR in handling the quantum corrections in a curved background. Therefore, using DR in a curved background, the scale and trace anomalies can both be obtained by counterterms that are Weyl invariant only at $d=4$. 
\end{abstract}

\maketitle


\section{Introduction}
The integration of a conformal sector in the early universe, is expected to carry significant implications for its cosmological evolution, since it is responsible for the generation of corrections to Einstein's gravity which take a very special form. These are quadratic in the Riemann tensor and are associated with just two specific Weyl invariant contributions, one of them being topological, the 
Gauss-Bonnet (GB) ($\sqrt{g} E$ term), while the second one is given by the Weyl tensor squared term ($C^2$). Conformal gravity, for example, is entirely based on such second invariant at $d=4$. These two terms are the only ones which appear to be necessary, in order to regulate the quantum corrections coming from a conformal sector, at least around flat 
space, and to generate the trace anomaly \cite{Duff:1993wm,Duff:1977ay,Capper:1974ic} in the presence of stress energy tensors. Correlators of this form are at the center of our current analysis.\\
Perturbative computations of conformal invariant sectors in external metric backgrounds are characterised by two essential features: 1) the presence of a logarithmic contribution involving a renormalization scale $(\mu)$, carrying the signature of broken scale invariance, and 2) pole terms in the correlation functions associated with the trace anomaly \cite{Giannotti:2008cv} \cite{Armillis:2009pq, Coriano:2018zdo}. Polar contributions are combined with traceless sectors in order to define a decomposition of a generic correlation function in terms of two parts, each of them conserved and satisfying, respectively, ordinary and anomalous conformal Ward identities \cite{Coriano:2021nvn,Coriano:2022jkn}. 
The full implications of the structure of the hierarchy of such constraints starts to emerge at the level of 4-point 
functions, treated by extending the reconstruction method of \cite{Bzowski:2013sza}, formulated for 3-point functions, to higher point functions in momentum space \cite{Coriano:2022jkn}.\\ 
Such logarithmic and non logarithmic contributions, according to a classification that we are going to critically review, have been at the center of several analysis. In particular, one of the standing issues is whether logarithmic corrections may result from the expansion around flat space of conformally invariant operators in $d$ dimensions generating new nonlocal conformal anomalies \cite{Deser:1976yx}.\\
As shown in \cite{Coriano:2017mux,Coriano:2021nvn}, the analysis of the effective action and of its anomalous Weyl variation, manifest in the form of constraints in tangent (Minkowski) space, directly linked with the CWIs of the correlation functions, by a functional expansion. This allows to address the consistency of the effective action by a study of the correlation functions and of their conformal hierarchies directly from momentum space \cite{Coriano:2013jba,Bzowski:2015pba}.  \\
Such constraints are deemed to be necessary in order to provide a consistent regularization of the theory in $d$ dimensions. It is quite clear that if such terms induce a scale dependence, naturally associated with a $\log$ contribution, this is attributed to the renormalization procedure, as one takes the flat spacetime limit. \\
One approach consists in using dimensional regularization (DR), formally extended to a curved spacetime, in such a way that, for a flat metric, one recovers the usual version of this method in Minkowski space. However, it is by no means guaranteed that this procedure is the most accurate one in the extraction of the anomaly content of a certain theory. This is expecially true if one relies on the assumption that the counterterms introduced in the regularization procedure, should be manifestly Weyl invariant for general spacetime dimensions, and not just at $d=4$.\\
 While the requirement that the counterterms respect the fundamental symmetry of the action is a reasonable assumption in flat space, in a general metric background this procedure is essentially ambiguous, at least in DR, 
since it requires the introduction of nonlocal operators  whose covariant expressions have so far not been identified, nor their flat spacetime limit is guaranteed to exist. This point has been recently emphasized in \cite{Coriano:2022jkn}.\\
 The background also introduces extra degrees of freedom of the metric as part of the effective action, due to the presence of extra dimensions. In addition, one has to perform a dimensional reduction (DRed) of the coordinate dependence of the fields, which is naturally associated with integrations regulated by extra cutoffs. \\
In this note we are going to investigate this issue, that illustrates the clear limitations of DR in the identification of the effective action for a curved spacetime. At the same time, as we are going to show, the same regularization procedure correctly accounts for all the anomalies (scale and trace anomalies) rather naturally in $d=4$, once the (local) Weyl invariant version of such counterterms are used for the regularization of the quantum corrections.\\
 In this work, anomalies of type "A" and "B", in the classification of \cite{Deser:1993yx}, with or without the generation of scale-breaking $\log$ contributions, are naturally accounted for by an appropriate subtraction scheme that does not require any nonlocal operator. The regularization procedure of the effective action that we introduce, differs by finite terms from the usual Wess-Zumino form. The extra terms involved in this regularization are Weyl invariant and, at the same time, include a log correction only for the 
$C^2$ counterterm. This defines at $d=4$ an anomaly of "B" type. For the topological contribution such a log is absent, since it amounts just to a constant added to the anomaly effective action, as expected for an anomaly of type "A".

\section{Weyl invariance and DR}
We start with some observations that can help to clarify the point of our discussion. \\
Given a functional $ F(g) $ of a certain metric $ g $, its invariance under a Weyl transformation requires that
after a redefinition of the metric of the form
\beq
\label{dc}
g_{\mu\nu}=\bar{g}_{\mu\nu} e^{2 \phi},
\eeq
where $\bar{g}$ is the fiducial metric and $\phi$ is the conformal factor, 
the functional $F$ remains invariant, that is 
\beq
\label{ss}
F(g)= F(\bar g).
\eeq
Eq. \eqref{dc} defines a conformal decomposition of the metric $g$, which is left invariant by the ($\sigma$) symmetry
 
\beqa
\label{sigma}
\phi\to \phi + \sigma \nn\\
\bar g_{\mu\nu}\to \bar g_{\mu\nu}e^{2 \sigma}.
\eeqa
A Weyl variation of the metric is simply associated with the active transformation 
\beq
g_{\mu\nu}\to g_{\mu\nu}e^{2\sigma}
\eeq 
for a generic local function $\sigma(x)$.
Notice that in the presence of a renormalization procedure such as DR, both conditions \eqref{dc} and \eqref{sigma} 
are violated, as we are going to elaborate below. On the other end, a functional may be Weyl non-invariant and can still satisfy the $\sigma$ symmetry. For instance, counterterms $V_E, V_{C^2}$ and $V_{F^2}$ 
\begin{align}
\label{ffr}
V_{C^2}(g, d)\equiv & \mu^{\epsilon}\int\,d^dx\,\sqrt{-g}\, C^2, \notag \\
V_{E}(g,d)\equiv &\mu^{\epsilon} \int\,d^dx\,\sqrt{-g}\,E  \notag \\
V_{F^2}(g,d)\equiv &\mu^{\epsilon} \int\,d^dx\,\sqrt{-g}\,F^2 
\end{align}
where 
\beqa
\label{GB1}
 E& =& R^2 - 4 R^{\mu \nu} R_{\mu \nu} + R^{\mu \nu \rho \sigma} R_{\mu \nu \rho \sigma},
\eeqa
is the Euler-Poincar\`e density 
and 
\beqa
\label{Geometry1}
C^{(d) \alpha\beta\gamma\delta}C^{(d)}_{\alpha\beta\gamma\delta}
&=&
R^{\alpha\beta\gamma\delta}R_{\alpha\beta\gamma\delta} -\frac{4}{d-2}R^{\alpha\beta}R_{\alpha\beta}\nonumber \\
&& +\frac{2}{(d-2)(d-1)}R^2
\eeqa
is the Weyl tensor squared. They are such that for $d\neq 4$ 

\beq
\label{ffr1}
2 g_{\mu\nu}\frac{\delta}{\delta g_{\mu\nu}}V_E=\epsilon \sqrt{g} E, 
\eeq 
and similarlyfor $V_{C^2}$ and $V_{F^2}$, with $F_{\mu\nu}$ being the field strenght of a spin-1.
They identify components of the stress energy tensor which are Weyl variant at $d=4$.  However, they still respect the $\sigma$ symmetry for being functionals of the complete metric $g$. 
In the expressions above, variations of the metric induced by varying the conformal factor $\phi(x)$ are obtained using the rule 
\beq
\label{var}
\frac{\delta }{\delta \phi}=2 g_{\mu\nu}\frac{\delta}{\delta g_{\mu\nu}},
\eeq
where the fiducial metric is kept fixed. Once the renormalization procedure has singled out a specific fiducial metric 
and a dilaton field, more general variations are possible, which involve independently $\bar{g}$ and $\phi$, whose dynamics is constrained by the anomaly. In this case one has to be careful with the use of \eqref{var}, since in  the regularization terms of $O(\epsilon) $ are dropped, and the $\delta_\sigma$ variation of \eqref{sigma} is nonzero.\\
 This procedure lays at the core of the derivation of the usual Wess Zumino (WZ) effective action, in the conformal anomaly context. \\
The WZ action is generated by subtracting two functionals, one dependent on $g$ and another one on $\bar{g}$, that we will call 
$\sm_R$ in our case, the renormalized effective action. It can be derived in DR, but subject to ambiguities which are typical of this scheme in a curved background, since the regularization is extended, with some difficulty, from the flat case.
For example, the traceless (Weyl invariant) contributions, which are part of the effective action, are ambiguous in this scheme, if one follows a straightforward renormalization procedure adapted from flat space \cite{Coriano:2022ftl}. The result of this procedure is to identify   
the fiducial metric and the dilaton $\phi$ as independent components, that need to be varied independently in the derivation of their equations of motion. It is also possible, in some cases, to remove the dilaton from the spectrum, 
by choosing a specific (integrable) conformal decomposition (see for instance \cite{Barvinsky:1995it}). 
In general this requires, from the point of view of DR, a finite renormalization of the topological term $V _E$ \cite{Mazur:2001aa}. A recent study of that includes also a dilaton in the background, beside the metric, is in  \cite{Asorey:2022ebz}.  \\
 As discussed in \cite{Coriano:2022ftl}, the renormalized action can take several forms, depending on the way we introduce the subtractions. 
 As just mentioned, the common one is to perform a subtraction using the fiducial metric as a reference, which has a well-defined meaning in the context of a renormalization procedure, but it is not the only possible one. One can show that if the WZ action is interpreted as the result of an ordinary regularization, implemented quite similarly to flat space, there are no logs generated in the Weyl variant part of the action. The reason, as we are going to clarify, is that this regularization misses some important Weyl invariant terms. These are the terms that, once the DRed procedure is kept into account, become responsible for the breaking of scale invariance.\\
As we are going to show, these appear in a natural way if the $1/(d-4)$ pole subtraction, in the Weyl invariant contributions - hereinafter referred to as $\sm_B$ - is implemented with respect to the same metric $g$ instead of $\bar{g}$, and directly at $d=4$. 
This subtle point can easily be recognized by the Weyl scaling expression of the counterterms, if one does not ignore the extra dependencies of the field coordinates, with the generation of natural cutoffs in the integration domain of the fields. 

\section{Review of regularization procedures} 
 We are going to review the regularization procedure to clarify this point. 
 
For this purpose, one considers the bare effective action 
 \beq
\label{deff}
e^{-\sm_B(g)}=\int D\chi e^{-\sm_0(\chi,g)},
 \eeq
and assume that $\sm_0(\chi,g)$, the classical action, is conformally invariant 
\beq
\sm_0(e^{-\Delta \sigma(x)}\chi,e^{2\sigma}g)=\sm_0(\chi,g), 
\eeq
with $\Delta$ denoting the scaling dimension of $\chi$. 
The singularities of the perturbative expansion 
will be encountered as $d\to 4$, and can be removed, for general backgrounds, by the inclusion of three counterterms. Beside $E$ and $C^2$, also the square of the field strength $F^2$ is needed. An example is provided by a non-vanishing spin-1 external field. In this case the renormalized hierarchy of the CWIs, for instance for the $TJJ$ - with a stress energy tensor $T$ and two spin-1 currents $J$ - will have a renormalized trace Ward identity only with the inclusion of the $F^2$ counterterm, since this projects on the $JJ$ 2-point function.\\
From now on we will consider only a metric $g$ in the background and exclude external photons. 
In this case, the renormalized action $\sm_R$ is constructed with the inclusion of counterterms of the form $(1/\epsilon) V$, using, in this case, only two of the three expressions in \eqref{ffr} 

\begin{equation}
e^{-\sm_R(g)}=\, {N}\int D\chi e^{-S_0(g,\Phi) + b' \frac{1}{\epsilon}V_E(g,d) + b \frac{1}{\epsilon}V_{C^2}(g,d) },
\end{equation} 
  where ${N}$ is a normalization constant. Here, $b,$ $b'$ count the number of massless fields involved in the loop corrections. The counterterms remove the $1/\epsilon$ singularities present in the bare effective action $\sm_B$
\begin{equation}
\sm_B(g,d)=-\log\left(\int D\chi e^{-S_0(\chi,g)}\right) +\log {N},
\end{equation}
as $d\to 4$. $\sm_B$ generates finite correlators in general dimensions.\\
The inclusion of non conformal sectors in the loop corrections, in a perturbative evaluation $\sm_B(g,d)$, will be addressed in a final section section, when we will discuss the case of the 2-point function of the $TT$ correlator, with the exchange of spin-1 matter in the loops. Also in that case we are going to show the consistency of the DR procedure with the inclusion of only local counterterms.

\subsection{Finite subtractions}
If we use DR as a regularization procedure, we need to face the ambiguity inherent in the choice of the finite subtractions, which are typical of any regularization scheme. We are going to see that such subtractions can be responsible for the generation of a scale dependence in the renormalized effective action, without the need of introducing nonlocal extensions of the counterterms for $d\neq 4$. We will argue that if the anomaly is exclusively a 4-dimensional phenomenon, such nonlocal terms appear to be shortcomings of DR rather than consistent predictions of the regulated theory. Let us now go into detail on this point. \\
In general, in the definition of a WZ action, the subtraction of the singularities present in $\sm_B$ generated as $d\to 4$, is performed with respect to a fiducial metric $\bar{g}$. But we may follow a slightly different approach, using only one metric, quite closely to how DR is implemented in flat space. First, we introduce the counterterm action 

 \begin{equation}
\sm_{v}\equiv \frac{b'}{\epsilon}V_E(g,d) + \frac{b}{\epsilon}V_{C^2}(g,d),
\end{equation}

and define the renormalized action $\sm_R$
 \begin{equation}
\label{decomp2}
\sm_R=\sm_B +\sm_v,
\end{equation}
and expand the counterterms respect to the fiducial metric 

\beqa
\sm_R(4)&=&\lim_{d\to 4}\left(\sm_B(g,d) +\frac{b'}{\epsilon}V_E(g,d) + \frac{b}{\epsilon}V_{C^2}(g,d)\right. \nonumber \\
&=&\sm_f +b' \,V'_E(\bar g,\phi, 4) + b\,  V'_{C^2}(\bar g,\phi, 4)\nonumber \\
\label{ren1}
\eeqa
where $\sm_f$ is finite and 
\beqa
\label{ps}
 V'(\bar g,\phi,4)&\equiv& \lim_{d\to 4}\left(\frac{1}{\epsilon}\left(V(g, d)- V(g, 4)\right)\right).\nonumber \\
\eeqa
Notice that the subtraction is performed respect to the same metric $g$, but the residue is expanded, according with to the DR approach in flat space, {\em directly at $d=4$}. One can easily show, however, that the subtraction can be equally expressed in terms of the fiducial metric $\bar{g}$, as far as we stay at $d=4$, since 
 \beq
 V(\bar g, 4)= V( g, 4) +\textrm{boundary terms},
 \eeq 
 that can be easily derived from the scaling relation shown in \eqref{form1}.\\
 As pointed out in \cite{Coriano:2022ftl}, the subtraction may involve $V_E(\bar g,4)$, but may as well involve $V_E(\bar g,d)$ with 
$d\to 4$, since DR may contemplate both cases. In this second case one replaces \eqref{ps} with 

\beqa
\label{ps1}
\hat V'_E&=& \lim_{d\to 4}\left(\frac{1}{\epsilon}\left(V_E(g, d)- V_E(\bar g, d)\right)\right),
\eeqa
with a renormalized effective action now rearranged in the form 
\beqa
\label{rem}
\sm_R&=&\tilde\sm_f(4) +b' \,\hat V'_E(\bar g,\phi, 4) + b\, \hat V'_{C^2}(\bar g,\phi, 4).\nonumber \\
\eeqa
A similar expression is obtained by using \eqref{ps} 
\beqa
\label{rem}
\sm_R&=&\sm_f(4) +b' \,V'_E(\bar g,\phi, 4) + b\, V'_{C^2}(\bar g,\phi, 4).\nonumber \\
\eeqa
Notice that both $\tilde\sm_f(4)$ and $\sm_f(4)$, that will be detailed below, identify finite functionals whose expansions in the background metric generate finite correlation functions. They will satisfy ordinary (non anomalous) hierarchies of CWIs. 
On the other end, the remaining ($V', \hat{V}'$) terms generate, by functional differentiations respect to the external metric, satisfy hierarchies of anomalous CWIs. \\
For instance, the anomalous variation of $\sm_R$ in \eqref{rem}, is entirely generated, in this regularization, by the remaining terms of 
\eqref{rem} 
\begin{equation}
\label{SA}
\sm_A\equiv \sm_{WZ}=b' \,\hat V'_{E}(\bar g,\phi,4) + b\,\hat V'_{C^2}(\bar g,\phi,4) + c \,\hat V'_{F^2}(\bar g,\phi,4),
\end{equation}
that takes a WZ form. We can conveniently rewrite it as
\beqa
\label{WWZZ}
S_{WZ}(g)&=& \lim_{d\to 4}\mu^\epsilon\left(\frac{\int d^d x \sqrt{g}\left(b' E + b C^2 + c F^2\right)}{\epsilon}\right.\nonumber \\
&& \left. - \int d^d x \frac{\sqrt{\bar g}\left(b' \bar E + b \bar C^2 + c \bar F^2\right) }{\epsilon}\right).
\eeqa
The barred invariants are all computed in the $\bar{g}$ fiducial metric. 
The $\phi$ dependence of the previous equation becomes explicit by the scaling relations 
\beqa\label{form1}
 E&=&\sqrt{\bar g} e^{(d-4)\phi}\biggl \{ \bar E+(d-3)\bar\nabla_\mu \bar J^\mu(\bar{g},\phi)\nonumber \\
 &&  +(d-3)(d-4)\bar  K(\bar{g},\phi)  \biggl \},
\eeqa
where we have defined
\beqa \label{GBexJ}
\bar J^\mu(\bar{g},\phi)&=& 8\bar R^{\mu\nu}\bar\nabla_\nu\phi -4\bar R\bar \nabla^\mu\phi  -4(d-2)(\bar\nabla^\mu\phi\bar \Box \phi- \nonumber \\
&&\bar \nabla^\mu\bar\nabla^\nu\phi\bar \nabla_\nu\phi+\bar\nabla^\mu\phi\bar\nabla_\lambda\phi\bar\nabla^\lambda\phi),
\eeqa
and
\beqa \label{GBexK}
 \bar K(\bar{g},\phi)&=& 4\bar R^{\mu\nu}\bar\nabla_\mu\phi\bar\nabla_\nu\phi-2\bar R\bar\nabla_\lambda\phi\bar\nabla^\lambda\phi \nonumber \\
&& +4(d-2)\bar\Box\phi\bar\nabla_\lambda\phi\bar\nabla^\lambda\phi+(d-1)(d-2)(\bnabla_\lambda \phi\bnabla^\lambda \phi)^2.\nonumber \\
\eeqa
We recall that the Weyl scaling relation 
\beq
\sqrt{g}C^2=e^{\epsilon \phi}\sqrt{\bar g} \bar C^2,
\eeq
with a similar rescaling being valid for $F^2$
\beq
\sqrt{g}F^2=e^{\epsilon \phi}\sqrt{\bar g} \bar F^2,
\eeq
require ad-hoc nonlocal modifications of such terms in order to respect Weyl invariance for a generic spacetime dimension $d$. The scaling provides the most direct way to extract the anomaly induced action in general dimensions \cite{Matsumoto:2022fln}\cite{Coriano:2013nja}\cite{Ferreira:2017wqz}\cite{Elvang:2012st}\cite{Komargodski:2011vj}.

\section{The WZ action and log terms}

The WZ action, viewed as a regularization procedure, is just one of the possible options in the definition of $\sm_R$, since Weyl invariant terms are obviously missing from this action. Indeed any regularization procedure has its own limitations.\\
 In the case of the conformal anomaly  - differently  from the chiral anomaly - 
the breaking of conformal symmetry is not entirely accounted for by a topological contribution. For example, 
there are no external Ward identities that one can impose in order to remove the ambiguities of a renormalization scheme due to the inclusion of finite terms. This is, instead, possible for an AVV chiral anomaly diagram.  \\
A subtle point to be taken into consideration in the presence of non topological contributions to the anomaly, is the emergence of possible conflicts between the conservation WIs and the choice of the finite subtraction terms in the effective action. After a functional expansion of the renormalized action, such terms may jeopardize the conservation WIs. For this reason we proceed with few simple considerations, linking the behaviour of DR in flat and curved spacetimes.\\
The analysis of the renormalization procedure in DR is quite instructive if applied to the flat case, and provides a well-defined example for the  analysis of correlators of stress energy tensors. For this reason, the subtraction of the singularity in $\sm_B$ is performed at $d=4$  according to \eqref{ps}, with a background Euclidean metric $g\to \delta$. 
As we are going to see, such subtractions preserve the conservation Ward identities, for being proportional to $\sqrt{g}C^2$ (times a scale breaking log).  As already mentioned, \eqref{ps1} modifies this subtraction by extra terms that we are going to identify. 
\subsection{WZ and other subtractions}
The renormalization procedure has allowed us to identify both $\phi$ and $\bar{g}$ as independent components, or, equivalently, $g$ and $\bar g$. \\
As already pointed out, we can immediately write down the complete, renormalized effective action in the form 
\beq
\label{tot}
\sm_R=\tilde\sm_f + S_{WZ}(\bar g,\phi)
\eeq
as sum of a contribution which is Weyl invariant $\tilde \sm_f$, and of a second term which is responsible for the generation of an anomaly. It is easy to figure out that $\tilde \sm_f$ is indeed Weyl invariant ($\delta_\phi \tilde \sm_f=0$), 
since 
\beq
\tilde \sm_f=\lim_{d\to 4}\left(\sm_B +b'\frac{1}{\epsilon}V_{C^2}(\bar g,d) + b \frac{1}{\epsilon}V_{E}(\bar g,d)\right) 
\eeq
is the sum of a Weyl invariant bare action $\sm_B$, and of two counterterms which are $\phi$ invariant for being functionals of the background metric $\bar{g}$. Here we will are assuming that the $d\to 4$ limit and the $\delta_\phi$ variation commute. \\
A similar result holds if we define the finite action $\sm_f$ using the subtraction \eqref{ps}, which is close to the usual DR renormalization in flat space since 
\beq
\sm_f=\lim_{d\to 4}\left(\sm_B +b'\frac{1}{\epsilon}V_{C^2}(g,4) + b \frac{1}{\epsilon}V_{E}(g,4)\right). 
\eeq
Both $\sm_f$ and $\tilde\sm_f$ are Weyl invariant. 
 Notice that from \eqref{tot}, if we send $g\to \bar g$, then 
\beq
\sm_{WZ}(\bar g)=0
\eeq
giving 
\beq
\sm_R(\bar g)=\tilde \sm_f. 
\eeq
In this formalism we recover the standard relation 
\beq
\label{ttwo}
\sm_R(g)-\sm_R(\bar g)= \sm_{WZ}, 
\eeq
which is usually derived directly from the definition of  $\sm_R$ via the path integral, formulated in $d$ dimensions. 
The two regularizations hide logs contributions in a subtle way since the two anomalous contributions 
$V'$ and $\hat V'$ differ by finite terms. For the Euler-Poincar\`e counterterm  
\beqa
\label{wwzp}
\hat{V}'_E(\bar g,\phi)&\equiv&\lim_{d\to 4}\left(\frac{1}{\epsilon}\left(V_E(g,d)-V_E(\bar g, d)\right)\right)\nonumber \\
&&=V'_E- \frac{\partial}{\partial d}V_E(\bar g, d)\mid_{d=4},
\eeqa
 and similarly for the other counterterms.
\eqref{wwzp} is a relation that can be made explicit by using the rescaling of the GB term in $d$ dimensions

\beq
\label{wzp2}
\hat{V}'_E(\bar g,\phi)=\int d^d x \sqrt{\bar g} \phi\left( \bar E + \bar\nabla_\mu \bar J^\mu \right) 
+\int d^d x \sqrt{\bar g} K.
\eeq
Therefore, the complete contribution to $\sm_A$ in \eqref{SA} coming from the $V_E$ counterterm, is given by 
\beqa
\label{like}
V'_E&\equiv&\lim_{d\to 4}\frac{1}{\epsilon}\left( V_E(g,d)-V_E(g,4)\right)=
\frac{\partial}{\partial d}V_E(\bar g, d)\mid_{d=4} \nonumber \\
&& +
\int d^4 x \sqrt{\bar g} \phi\left( \bar E + \bar\nabla_M \bar J^M \right) 
+\int d^4 x \sqrt{\bar g} K,
\eeqa	
where the first terms on the rhs is due to the difference 
\beq
\frac{\partial}{\partial d}V_E(\bar g, d)\mid_{d=4}=\lim_{d\to 4}\frac{1}{\epsilon}\left( \int d^d x \sqrt{\bar{g}}\bar E - \int d^4 x \sqrt{\bar{g}}\bar E\right).
\label{difE}
\eeq
Similarly, for the $C^2$ counterterm we have  
\beqa
V'_{C^2}&\equiv& \lim_{d\to 4}\frac{1}{\epsilon}\left( V_{C^2}(g,d)-V_{C^2}( g,4)\right)\nonumber \\
&=&
\frac{\partial}{\partial d}V_{C^2}(\bar g, d)\mid_{d=4} + \int d^d x \sqrt{\bar g} \phi \bar C^2,
\eeqa
where 
\beq
\frac{\partial}{\partial d}V_{C^2}(\bar g, d)\mid_{d=4}= \lim_{d\to 4}\frac{1}{\epsilon}\left( \int d^d x \sqrt{\bar{g}}\bar C^2 - \int d^4 x \sqrt{\bar{g}}\bar C^2\right).
\label{difC}
\eeq
Simple manipulations show that $\hat{V}'$ and $V'$ differ by Weyl invariant  logarithmic terms, once we perform a dimensional reduction (DRed) of the rhs of \eqref{difC}. The cases of $V_E$ and of $V_{C^2}$, 
(or $V_{F^2}$) are, however, different, since $V_E$ is topological at $d=4$ and, indeed, plays a direct role as a finite renormalization of the effective action, while $V_{C^2}$ is necessary for the cancelation of the $1/\epsilon$ singularity. Both terms can be treated similarly. An expansion gives 

\bea \label{WG in generic case}
\frac{1}{d-4}V_E(g,d) =\frac{1}{\epsilon} \left({L}{\mu}\right)^{\epsilon}\int d^4x \rg  \bar{E}&&\nonumber \\
 + \left({L}{\mu}\right)^{\epsilon}\int d^4x \rg\ \Big[\phi {}_4\bar{ E}-(4 {} G^{\mu\nu}(\bar\nabla_\mu\phi\bar\nabla_\nu\phi) &&\nn \\
+2(\nabla_\lambda \phi \nabla^\lambda \phi )^2 +4\Box\phi \nabla_\lambda \phi \nabla^\lambda \phi ) \Big],
&&
\eea 
where the cutoff $L$ has been introduced here to bound the integrands over the extra dimensions. The expansion is accompanied by the ordinary renormalization scale $\mu$.  It is easy to figure out, however, that due to the topological nature of the GB term at $d=4$, the term 
\beq
 \frac{1}{\epsilon} \left({L}{\mu}\right)^{\epsilon}\int d^4x \rg  \bar{E}=(\frac{1}{\epsilon}+ \log(\mu L) )\int d^4x \rg  \bar{E} +O(\epsilon)
\eeq
 does not contribute to the equations of motion. A similar results holds also for the  $V_{C^2}$, which is part of the WZ action. Even if not topological, also this term will not contribute to the generation of scale variant contributions, due to the cancelations of the log terms coming from the expansion of $\hat V'_{C^2}$ 
\beqa
\hat{V}'_{C^2}&=& \frac{1}{\epsilon}\left( 1 +\epsilon \log \mu L\right) \int d^4 x \sqrt{\bar g}(1 + \epsilon \phi)\bar C^2 \nonumber \\
&& - \frac{1}{\epsilon}\left( 1 +\epsilon \log \mu L\right) \int d^4 x \sqrt{\bar g}\bar C^2.
\eeqa 
It is clear from the expression above that the cancelation of the log terms is natural, at least if we follow this prescription, as we perform the $\epsilon\to 0$ limit.  \\
The pattern changes drastically if we perform subtractions defined by the $V'$ terms, as in \eqref{difC} that, instead, generate contributions of the form 
\beq
\label{term}
\frac{\partial}{\partial d}V_{C^2}(\bar g, d)\mid_{d=4}=\log\mu L \int d^4 x \sqrt{\bar g}\bar C^2.
\eeq 
A similar results holds for the ${V_{F^2}}$ terms in the presence of a spin-1 background. \\
As already mentioned, this term breaks the dilatation symmetry and is Weyl invariant. It is also quite clear that even with the choice of a minimal subtraction $V_E(\bar g, 4)$ instead of a modified one ($V_E(\bar g,d)$), if we proceed with DRed on the metric, the term analogous to \eqref{term} would be 
\beq
\label{term}
\frac{\partial}{\partial d}V_{E}(\bar g, d)\mid_{d=4}=\log\mu L \int d^4 x \sqrt{\bar g}\bar E,
\eeq 
which is topological and Weyl invariant, and therefore irrelevant in the effective action. \\
Therefore, we have shown 
that the scale dependence generated by the expansion of local counterterms,  is only linked to their topological or non-topological behaviour, if we follow closely a renormalization scheme such as DR, where we perform a subtraction scheme borrowed from flat space. 
 
 \section{Wess-Zumino Consistency conditions and renormalization}
In this section we are going to compare the approach followed in the definition of the renormalized effective action $\sm_R$ outlined in the previous sections, and the Fujikawa method, based in the inclusion of a cutoff $(M)$ at $d=4$. The use of this scheme does not allow to set a distinction between topological and non topological contributions as in DR. However, it is useful in order to show that the WZ consistency condition, that can be derived quite easily from the anomalous variation of the integration measure in the path integral, hides a log of the cutoff if we trace back all the steps of the derivation.\\
We redefine the metric according to \eqref{dc} and perform the change in the measure 
\beq
\label{anomvar}
D\chi\vert_{\bar g e^{2 \phi}}=D\chi\vert_{\bar g}e^{-\sm_{p}(\phi, \bar g)},
\eeq
 where we have indicated with $\sm_{p}$ the action generated by the change of variables in the metric. In the Minkowski case this corresponds to a phase coming from the anomalous variation, first computed in \cite{Fujikawa:1980rc,Fujikawa:1980vr}. This result holds only if we perform such variation in four dimensions and it requires a cutoff regularization. \\
 Alternatively, for a sector which is conformal in $d$ dimensions, if we use DR, we will attribute the anomaly to the 
 counterterms $V_E$ and $V_{C^2} $ rather than to the anomalous variation of the integration measure, as clear from the anomaly constraints \eqref{ffr}.\\
We start to address this point by assuming the validity of \eqref{anomvar}. We keep implicit the cutoff dependence of this equation. The anomalous variation will then be reconsidered afterwards at a finer level, following closely the steps of \cite{Fujikawa:1980rc,Fujikawa:1980vr}, by taking explicitly into account the cutoff regularization. 
 
The effective action has been defined in \eqref{deff}
 where we assume that $\sm_B(g)$ is Weyl invariant in $d$ dimensions.  If the sector $\chi$ is conformal, than 
 clearly $\sm_B(g)=\sm_B(\bar g)$ and $\sm_0(\chi,g)=\sm_0(\chi,\bar g)$ for any conformal decomposition.\\
The usual approach to derive the WZ consistency condition is to require that the anomalous variation of the functional integral comes from the measure of integration, as shown in \eqref{anomvar} as we select $d=4$. Then, on the rhs of \eqref{deff} we introduce the conformal decomposition \eqref{dc} to obtain 

\beqa
\label{bbare}
e^{-\sm_B(g)}&=& \int D\chi\vert_{\bar g}e^{-\sm_{p}(\phi, \bar g)}e^{-\sm_0(\chi,\bar g)} \nn\\
&=&  e^{-\sm_{p}(\phi, \bar g)} \int D\chi\vert_{\bar g}e^{-\sm_0(\chi,\bar g)}\nn\\
&=& e^{-\sm_{p}(\phi,\bar g)} e^{-\sm_B(\bar g)},
\eeqa
 with $\sm_B(g)\neq \sm_B(\bar g)$. The difference between the two actions is related to the "phase contribution" 
$\sm_p$, giving 
\beq
\label{one}
\sm_B(g) - \sm_B(\bar g)=\sm_{p}(\phi,\bar g).
\eeq
At this point, we require that $\sm_B(g)$ is a functional which is invariant under the $\delta_\sigma$ variation \eqref{sigma} and use again \eqref{anomvar} with $\phi\to \sigma$ to obtain
\beqa
e^{-\sm_B(g)}&=&\int D\chi\vert_{\bar g} e^{-\sm_{p}(\sigma, \bar g)} e^{-\sm_{p}(\phi-\sigma,\bar g e^{2 \sigma})}
e^{-\sm_0(\chi,\bar g)} \nonumber \\
& =& e^{-\sm_{p}(\sigma, \bar g)} e^{-\sm_{p}(\phi-\sigma,\bar g e^{2 \sigma})} e^{-\sm_B(\bar g)},
\eeqa
and
\beq
\label{onepp}
\sm_B(g) - \sm_B(\bar g)= \sm_{p}(\sigma, \bar g) + \sm_{p}(\phi-\sigma,\bar g e^{2 \sigma}).
\eeq
This equation can be combined with \eqref{one} to generate the WZ consistency condition
\beq
\label{twopp}
\sm_{p}(\phi,\bar g)= \sm_{p}(\sigma, \bar g) + \sm_{p}(\phi-\sigma,\bar g e^{2 \sigma}).
\eeq
Now we are going to repeat the steps above but taking care of the renormalization procedure.\\
As we have already mentioned, we use a cutoff regularization and rely on the analysis of \cite{Fujikawa:1980rc,Fujikawa:1980vr}. 
For example, in the simple case of a conformal scalar, with the action 
 \beq
 \sm_{scalar}=\int d^4 x \sqrt{g}\left(\frac{1}{2}\partial_\mu\tilde\chi\partial^\mu\tilde\chi + 
 \frac{1}{12}R{\tilde\chi}^2\right),
 \eeq 
 the computation of the anomalous variation of the measure is performed by an expansion of the scalar field
 \beq
\tilde\chi{(x)}=\sum_n a_n \tilde\chi_n(x)
\eeq
in terms of eigenfunctions of the operator
\beq
\left(\Box -\frac{R}{6}\right)\tilde{\chi}_n(x)=\lambda_n\tilde\chi_n(x),
\eeq
followed by a four-dimensional cutoff regularization of the infinite sum 
\beq
\mathcal{A}=\sum_n\tilde\chi_n(x)\tilde\chi_n(x)\to \sum_n\tilde\chi_n(x)e^{-\lambda_n /M^2}\tilde\chi_n(x),
\eeq 
with $\mathcal{A}$ being the conformal anomaly contribution

\beq
g^{\mu\nu}\langle T_{\mu\nu}\rangle= - \mathcal{A}.
\eeq
Then, the change in the integration measure can be rewritten in the form 
\beq
D\tilde\chi\vert_{\bar g e^{2 \phi}}=D\tilde\chi\vert_{\bar{g}} e^{ 2\int \sqrt{g}\phi (x) \bar{\mathcal{A}}_W d^4 x}.
\eeq
The regulator $M$ takes part in the variation of the bare effective action. This includes also the conformal anomaly contribution (here for a single conformal scalar field)
\beqa
\bar{\mathcal{A}}_W&=&\lim_{M\to \infty}\left( \frac{M^4}{(4 \pi)^2} +\frac{1}{2880 \pi^2}\left( (\bar R_{\mu\nu\alpha\beta})^2 
\right.\right.\nonumber \\
&&\left.\left. - 4 (\bar R_{\mu\nu})^2 +\bar \Box \bar R\right) \right).
\eeqa
Notice that the first ($M^4$) contribution can be removed by a vacuum renormalization. In a standard perturbative picture 
this corresponds to a tadpole diagram with a single external graviton. A contribution of this type is absent in DR in the flat spacetime limit, since it would correspond to a massless tadpole, but not in this scheme.  \\
Obviously, in the presence of a regularization, the two steps above in \eqref{one} and \eqref{onepp} need to be modified. We decompose the regulated anomalous variation in the form
\beq
\label{sst}
\sm_P(\phi,\bar{g},M)= \sm_P(\phi,\bar{g}) + \tilde\sm_P(\phi,\bar{g},M),
\eeq
where
\beq
\sm_P(\phi,\bar{g}) = 2\int d^4 x \sqrt{g} \phi(x)\mathcal{A}(x),
\eeq
is identical with the previous definition in \eqref{bbare} and
\beq
\mathcal{A}= \frac{1}{2880 \pi^2}\left( (\bar R_{\mu\nu\alpha\beta})^2 
- 4 (\bar R_{\mu\nu})^2 +\bar \Box \bar R\right),
\eeq
is the (finite) expression of the anomaly for a single conformally coupled scalar field. The new cutoff dependent part in \eqref{sst} is given by
\beq
 \tilde{\sm}_P(\phi,\bar{g},M)=\frac{2}{(4 \pi)^2}\int d^4 x \sqrt{\bar g} \phi(x){M^4},
\eeq
With these changes, eq. \eqref{one}, that takes to the WZ consistency condition  \eqref{twopp},  turns into
 
\beq
\label{three}
\sm_B(g) - \sm_B(\bar g)=\sm_{p}(\phi,\bar g,M).
\eeq
 Notice that \eqref{twopp} is broken by the regularization since the cutoff dependent term of $\sm_P$, does not satisfy the same condition 
 
\beq
\label{two}
\tilde\sm_{p}(\phi,\bar g,M)\neq \tilde\sm_{p}(\sigma, \bar g,M) + \tilde\sm_{p}(\phi-\sigma,\bar g e^{2 \sigma},M).
\eeq
This is the signature that one needs to perform a renormalization of the effective action $\sm_P$ in order to satisfy \eqref{twopp}. As already mentioned, the counterterm that removes the $\sqrt{\bar g}\phi M^4$ divergence is, obviously, a tadpole with one insertion of the trace of the stress energy tensor. This contribution corresponds to a cosmological constant term computed in a covariant approach \cite{Donoghue:2020hoh}. \\
One should keep in mind that even if we may safely neglect the cutoff dependence in \eqref{twopp}, 
the WZ consistency condition is the result of both \eqref{one} and \eqref{twopp}. Therefore, it is clear 
that if we solve Eq. \eqref{twopp}, ignoring the cutoff dependence introduced by \eqref{three}, we are omitting 
the dependence on $M$ that is generated once $\sm_B(g)$ is renormalized by the same cutoff. Only if we enforce an exact cancelation of the the $M^4$ divergence, a residual scale dependence can be ignored, otherwise it is quite obvious that a log will appear naturally in the procedure.\\
Therefore, a renormalization procedure, if taken into account properly, tells us that the anomaly induced action, even it may satisfy the WZ consistency condition as given by \eqref{two}, should still account for the presence of a cutoff that breaks the scale invariance of the bare effective action $\sm_B(g)$. \\
However, we have seen that the WZ consistency condition can be formulated in such a way that the cutoff dependence of the relation can be essentially ignored. For this reason, the analysis of such constraints can be consistently formulated as in \eqref{WWZZ} \cite{Antoniadis:1992xu}, and  the logarithmic terms can be ignored. 
In this approach, the scale-breaking contributions are just assumed to be part of the Weyl invariant sector of the effective action, but not of $\sm_{WZ}$ in \eqref{WWZZ}. 

\section{The issue of Weyl invariant counterterms in DR}
It is clear from our arguments, that the requirement of introducing Weyl invariant counterterms in DR for generic dimension $d$, motivated by the need to explain the dilatation anomaly, is questionable.\\
 As clear from \eqref{ffr} and \eqref{ffr1}, the structure of the $V_i(g,d)$ is such that at $d=4$ are Weyl invariant, ony at $d=4$. Weyl invariant extensions of such counterterms to $d$ dimensions have been motivated 
 in the approximate form \cite{Deser:1976yx}
\beq
\label{wr1}
\frac{1}{\epsilon}V_C^2\to \frac{1}{\epsilon}C_{\mu\nu\rho\sigma}\Box^\epsilon C^{\mu\nu\rho\sigma},
\eeq
and similarly for $V_{F^2}$. This would allow to regulate the theory using only counterterms which are Weyl invariant in $d$ dimensions, proceeding afterwords to $d=4$ by an expansion in $\epsilon$. The expansion of the $\Box^\epsilon$ term, in particular,  would generate both the ordinary counterterms $V_{C^2}$ and $V_{F^2}$ of 
$d=4$, and a much desired (scale breaking) log term from \eqref{wr1} as 
\beq
\label{wr2}
C\log \Box C\qquad \textrm{or}   \qquad F\log \Box F.
\eeq
A consistent definition of such  nonlocal extensions is still missing. \\
However, reasonable doubts exist which disprove the hypothesis that a non-local Weyl invariant regularization in dimension $d$ is necessary to recover the correct effective action. It should be clear, though, that an anomaly induced action, such as the Riegert \cite{Riegert:1984kt} action or the Fradkin-Vilkovisky action 
\cite{Fradkin:1978yw} \cite{Barvinsky:1995it} 
should not necessarily account for such contributions, being the aim of such actions only to reproduce the trace anomaly through a variational approach.
The inability of these actions to predict breaking of scale invariance is not an argument against their consistency. Other issues may be more important, to investigate their consistency, such as the absence of double poles around flat spacetime in the correlation functions computed from such functionals, or their consistency with perturbative calculations. These are separate issues, which are currently being investigated \cite{Coriano:2022jkn}. 
\subsection{Two examples}   
While it is not excluded that future progress towards a more general regularization scheme, such as an extension of DR, may be needed to connect expansions around flat backgrounds with those originating from curved spacetimes,
this need not change our current perspective on the origin of the conformal anomaly. This remains, from our point of view, a real $4d$ phenomenon.
 An important example comes from the analysis of correlators involving both conformal and non conformal sectors in their quantum corrections, using free field theory realizations. In general, at least for $d=2, 4$, the analysis is restricted to scalars, fermions and spin-1 fields running in the loops. Consider for example the 2-point function of stress energy tensors computed around flat space but in $d$ dimensions, with arbitrary numbers of scalars, fermions and spin-1 fields, with multiplicities $n_S, n_\psi,n_G$. In $d\neq 4$, if we compute this correlator in DR, we expect to find non conformal (Weyl variant) contributions, giving a nonvanishing trace, which is not anomaly related. A direct computation around flat space gives in $d$ dimensions 

\beqa
\braket{T^{\mu_1\nu_1}(p)T^{\mu_2\nu_2}(-p)}=-\frac{\p^2\,p^4}{4(d-1)(d+1)}\,B_0(p^2)\times&& \,\nonumber \\
\times \Pi^{\mu_1\nu_1\mu_2\nu_2}(p)\Big[2(d-1)n_F+(2d^2-3d-8)n_G+n_S\Big] +&& \nonumber \\ +
 \mathcal{N}^{\mu_1\nu_1\mu_2\nu_2}, &&\nonumber \\
\label{TTddim}
\eeqa
with
\beq
\label{N}
\mathcal{N}^{\mu_1\nu_1\mu_2\nu_2}=\frac{\p^2\,p^4\,n_G}{8(d-1)^2}(d-4)^2(d-2)\p^{\mu_1\nu_1}(p)\p^{\mu_2\nu_2}(p)\,B_0(p^2)
\eeq
where $B_0(p^2)$ is the scalar 2-point function, defined as
\begin{equation}
B_0(p^2)=\frac{1}{\pi^\frac{d}{2}}\,\int\,d^d\ell\,\frac{1}{\ell^2\,(\ell-p)^2}.
\end{equation}
There are some relevant features of this result, as we take the $d\to 4$ limit. The first contributions, proportional to the transverse traceless projector $\Pi$, is generated by all the three matter sectors, while the second term 
$\mathcal{N}$, proportional to $\epsilon ^2$ and $n_G$, contributes to the trace. The same term is only related to the presence of $n_G$ vector modes in the loop corrections. Indeed the spin-1 sector breaks Weyl invariance in $d\neq 4$, and for this reason we are dealing with a theory which is, overall, non conformal. \\
The fact that in $\mathcal{N}$ the prefactor is of $O(\epsilon^2)$ while all the contribution in $B_0(p^2)$ is of $O(1/\epsilon)$, is a clear indication that the breaking of the Weyl simmetry in $d\neq 4$ is explicit , since the other contributions, scalars and fermions, leave the correlator traceless. 
Notice, however, that this trace contribution vanishes as $\epsilon\to 0$. The result is suggestive of the fact that a non-conformal contribution for $d\neq 4$ in the action is not responsible for the generation either of a trace or a dilatation anomaly once we set $d=4$, since such non conformal behaviour is, after all, evanescent. \\
Since the regularization is taking care of the anomalous behaviour of the correlator - we are going to see this in a moment - even in  the non conformal case, it is unclear why Weyl invariant counterterms in $d$ dimensions are needed in the regularization of this correlator.\\
Let' s now come to discuss the renormalization of such correlator using a local counterterm. \\
The correlator is regulated only by the $C^2$ counterterm, since the second functional derivative of the $E$ counterterm vanishes 
\beq
V_{E}^{\mu_1\nu_1\mu_2\nu_2}(p,-p)=0.
\eeq
The counterterm takes the form of a transverse traceless projector around flat space
 \begin{align}
 \label{cc}
\braket{T^{\mu_1\nu_1}(p)T^{\mu_2\nu_2}(-p)}_{count}\equiv &\notag \\
-\sdfrac{\mu^{-\varepsilon}}{\varepsilon}\bigg(4b\,\big[\sqrt{-g}\,C^2\big]^{\mu_1\nu_1\mu_2\nu_2}(p,-p)\bigg)&\notag \\
=-\frac{8(d-3)\,\mu^{-\varepsilon}\,b}{(d-2)\,\varepsilon}p^4\Pi^{(d)\,\mu_1\nu_1\mu_2\nu_2}(p).
&\end{align}
The parameter $b$ is fixed by the cancellation of the singular $1/\epsilon$ behaviour, which in this case corresponds to the combination
\begin{equation}
\label{choiceparm1}
b=-\frac{3\pi^2}{720}n_S-\frac{9\pi^2}{360}n_F-\frac{18\pi^2}{360}n_G.
\end{equation}
As known, the choice of $C^2$ \eqref{Geometry1} with a parametric dependendence on $d$ which can be set to 4 either before or after the variation, allows to eliminate or keep the trace anomaly contribution in its second variation $\big[\sqrt{g}C^2\big]^{\mu_1\nu_1\mu_2\nu_2}$. \\
In both cases we are considering versions of the $V_{C^2}$ counterterm which have the same scaling behaviour under a Weyl transformation, but differing by a local renormalization. They generate different boundary contributions $\Box R$ after $\delta_\phi$ (Weyl) variations, and surely not a $\log \Box$ contribution to the anomaly.\\
The regularization of the $TT$ is implemented around $d=4$ by an expansion of the projector in \eqref{TTddim}
\beqa
\label{pexp}
\Pi^{\,\mu_1\nu_1\mu_2\nu_2}(p)&=&\Pi^{(4)\,\mu_1\nu_1\mu_2\nu_2}(p)-\frac{2}{9}\varepsilon\,\pi^{\mu_1\nu_1}(p)\,\pi^{\mu_2\nu_2}(p)\nonumber \\
&& +O(\varepsilon^2), 
\eeqa
and involves only the first term in \eqref{TTddim}, being the second term in the same equation, as remarked above, evanescent. \\
It is clear that as we perform an expansion of such projector around $d=4$ both in the bare $TT$
\eqref{TTddim} and in the counterterm \eqref{cc}, we obtain trace-free and trace contributions, with cancelations which are obviously unrelated to the $\mathcal{N}$ term. The local counterterm gives
 \begin{align}
\braket{T^{\mu_1\nu_1}(p)T^{\mu_2\nu_2}(-p)}_{count}=-\frac{8\,b\,p^4}{\varepsilon}\bigg(\Pi^{(4)\,\mu_1\nu_1\mu_2\nu_2}(p)&\notag \\
 -\frac{2}{9}\varepsilon\,\pi^{\mu_1\nu_1}(p)\,\pi^{\mu_2\nu_2}(p)+O(\varepsilon^2)\bigg)\times\notag \\
\times \bigg(\frac{1}{2}-\frac{\varepsilon}{2}\left(\frac{1}{2}+\log\mu\right)+O(\varepsilon^2)\bigg) && \notag \\
\hspace{1cm}=-\sdfrac{4\,b}{\varepsilon}p^4\,\Pi^{(4)\,\mu_1\nu_1\mu_2\nu_2}(p)\notag \\
+4\,b\, p^4\bigg[\Pi^{(4)\,\mu_1\nu_1\mu_2\nu_2}(p)+\frac{2}{9}\p^{\mu_1\nu_1}(p)\p^{\mu_2\nu_2}(p)\bigg]+O(\varepsilon)
&\end{align}
and the final renormalized expression becomes, using \eqref{choiceparm1}  for $b$ 
 \begin{align}
\braket{T^{\mu_1\nu_1}(p)T^{\mu_2\nu_2}(-p)}_{Ren}&=\braket{T^{\mu_1\nu_1}(p)T^{\mu_2\nu_2}(-p)}\notag\\
&+\braket{T^{\mu_1\nu_1}(p)T^{\mu_2\nu_2}(-p)}_{count}\notag\\
=-\frac{\p^2\,p^4}{60}\bar{B}_0\left(\frac{p^2}{\mu^2}\right)&\Pi^{\mu_1\nu_1\mu_2\nu_2}(p)\left(6n_F+12n_G+n_S\right)\notag\\
\quad-\frac{\p^2\,p^4}{900}\Pi^{\mu_1\nu_1\mu_2\nu_2}(p)&\big(126n_F-18n_G+31n_S\big).
\label{tren}
\end{align}
which is transverse and traceless. Henceforth, there is no trace anomaly. \\
Obviously, this result holds in the case in which we choose a counterterm 
$C^2$ parametrically dependent on $d$ $(i.e. (C^{(d)})^2)$.  If we had chosen its $4d$ version (i.e. with $d=4$) 
we would have found the relation 
\begin{equation}
\delta_{\mu_1\nu_1}\langle T^{\mu_1\nu_1}(p_1)T^{\mu_1\nu_2}(-p_1)\rangle=
\mathcal{A}^{\mu_2\nu_2}(p_1),
\end{equation}
where $\mathcal{A}^{\mu_2\nu_2}$ on the right hand is derived from the local  $\square R$ term of the anomaly. 
This is related to the identity
\begin{equation}
\frac{\delta}{(d-4)\delta \sigma (x) }\int d^d x \sqrt{-g} (C^{(4-\epsilon)})^2 =\sqrt{-g}\Bigg( (C^{(4)})^2
-\frac{2}{3}\square R\Bigg),
\label{twoq}
\end{equation}
that allows to get rid of the local (regularization dependent, $\Box R$) part of the anomaly, by a redefinition of the counterterm.
Notice that a dilatation anomaly is naturally present in the correlator since from \eqref{tren}
\begin{align}
\mu\frac{\partial}{\partial\mu}\braket{T^{\mu_1\nu_1}(p)T^{\mu_2\nu_2}(-p)}_{Ren}=&\notag\\
-2\left[\frac{\pi^2\,p^4}{60}\,\Pi^{\mu_1\nu_1\mu_2\nu_2}(p)\left(6n_F+12n_G+n_S\right)\right].&
\end{align}
This examples illustrates quite clearly that a non conformal theory in $d$ dimensions can be regulated in DR and generates the correct result for the expression of the correlator, just by the inclusion of a local counterterm. The regularization requires a subtraction which is Weyl invariant only at $d=4$. 
This result indicates that the anomaly is a genuine $4d$ phenomenon.\\
A similar analysis, obviously, can be carried out in $d=2$, where the log dependence is absent. In this case 
the anomaly is purely topological and given by the density $\sqrt{g} R$. Also in this case we consider both conformal and non conformal sectors in the quantum corrections. 
\subsection{Absence of scale breaking at $d=2$ in DR}
The behaviour of the 2-point function $TT$ in DR at $d=2$ is also quite illuminating and shows the consistency of the regularization procedure. In this case, for a conformal sector in the loops, we get 
 \begin{align}
	\braket{T^{\mu_1\nu_1}(p)T^{\mu_2\nu_2}(-p)}= \frac{ c(d)}{(d-2)}\left(p^2\right)^{d/2}\Pi^{\mu_1\nu_1\mu_2\nu_2}_{(d)}(p)\label{TT2dim}
\end{align}
where 
\begin{align}
	c(d)=4\,c_T \left(\frac{\pi}{4}\right)^{d/2}\frac{(d-1)\Gamma\left(2-\frac{d}{2}\right)}{\Gamma\left(d+2\right)}.
\end{align}
and
\begin{align}
	\braket{T^{\mu_1\nu_1}(p)T^{\mu_2\nu_2}(-p)}= \frac{ c(d)}{(d-2)}\left(p^2\right)^{d/2}\Pi^{\mu_1\nu_1\mu_2\nu_2}_{(d)}(p)\label{TT2dim}.
\end{align}

 $\Pi_{(d)}$ is, again, the transverse traceless projector defined in general $d$. The inclusion of a non conformal 
 sector is again handled by adding the same tensor $\mathcal{N}$ in \eqref{N}, whose expression is given for a generic $d$.  The correlator at $d=2$ does not require any counterterm to be regular. We can derive its 
trace anomaly, which is purely topological, either by closely investigating the degeneracy of its tensor structure as $d\to 2$ or, alternatively, by using a local counterterm. In both cases the result will be identical and, as foreseen from the "A", "B" classification of the anomaly, there will be no dilatation anomaly, being of type "A".\\  Note that the factor $1/(d-2)$ in \eqref{TT2dim} is purely kinematical, since the scalar loop appearing in the $TT$ is finite at $d=2$. One can easily show that as $d\to 2$, $\Pi^{\mu_1\nu_1\mu_2\nu_2}_{(d)}$ vanishes linearly in $(d-2)$, giving a finite expression for the $TT$. As mentioned, in the first approach we exploit the degeneration of tensor structures at $d=2$ for a $n$-point function - for $n=d$ - without the need to introduce a counterterm. This allows to re-express the Kronecker $\delta$ as 
 \begin{align}
 \label{dd}
	\delta^{\mu\nu}=\frac{1}{p^2}\left(p^\mu p^\nu+n^\mu n^\nu\right),
\end{align} 
with \begin{align}
	n^{\mu}=\epsilon^{\mu\nu}p_\nu,\label{indMom}
\end{align}
giving 
\begin{align}
	\Pi^{\mu_1\nu_1\mu_2\nu_2}_{d=2+2\epsilon}(p)&=\pi^{\mu_1(\mu_2}\pi^{\nu_2)\nu_1}-\frac{1}{1+2\epsilon}\pi^{\mu_1\nu_1}\pi^{\mu_2\nu_2}\notag\\
	&=\frac{2\epsilon}{1+2\epsilon}\frac{n^{\mu_1}n^{\nu_1}n^{\mu_2}n^{\nu_2}}{p^4},
\label{ths}
\end{align}
which is, as mentioned, of $O(d-2)$
while, right at $d=2$ the same projector vanishes if we use \eqref{dd}
\begin{equation}
	\Pi^{\mu_1\nu_1\mu_2\nu_2}_{(d=2)}(p)=\pi^{\mu_1(\mu_2}\pi^{\nu_2)\nu_1}-\pi^{\mu_1\nu_1}\pi^{\mu_2\nu_2}=0.
	\label{zerop}
\end{equation}
The result for the $TT$ at $d=2$ is 
\begin{align}
	\braket{T^{\mu_1\nu_1}(p)T^{\mu_2\nu_2}(-p)}_{(d=2)}= c(2)\frac{n^{\mu_1}n^{\nu_1}n^{\mu_2}n^{\nu_2}}{p^2}
\end{align} 
that reproduces the correct form of the anomaly in $d=2$. Indeed, by tracing the first two indices  one obtains
\begin{equation}
\label{ann}
	\braket{T^{\mu_1}_{\ \ \ \mu_1}(p)T^{\mu_2\nu_2}(-p)}_{(d=2)}^{Ren}=c(2)\,p^2\,\pi^{\mu_2\nu_2}(p).
\end{equation}
The right hand side of \eqref{ann} equals one functional derivative of trace anomaly of the 2-point function in two dimensions, in momentum space 
\beq
\langle T^\mu_\mu\rangle=\mathcal{A}_2
\eeq
with 
\begin{align}
	\mathcal{A}_2=c(2)\,\sqrt{-g}\,R
\end{align}  
being the anomaly at $d=2$. 

 The approach does not require any renormalization and henceforth no scale breaking log is generated. Thus, we have a trace anomaly but not a dilatation anomaly, as expected.
At this point, one may ask what kind of prediction follows from a DR regularization of the correlator, and whether the absence of a dilatation anomaly can be derived without resorting to the subtraction of non-local counterterms.\\

In this case, the inclusion of conformal sectors in the loops gives 
\begin{align}
	\braket{T^{\mu_1\nu_1}(p)T^{\mu_2\nu_2}(-p)}_{Reg}=& \frac{ c(2)}{2\varepsilon}\left(p^2\right)\,\Pi^{\mu_1\nu_1\mu_2\nu_2}_{(2+2\varepsilon)}(p)\notag \\
	+\frac{c(2)}{2}\Pi^{\mu_1\nu_1\mu_2\nu_2}_{(2)}\,p^2\log p^2 &+p^2c^\prime(2)\,\Pi^{\mu_1\nu_1\mu_2\nu_2}_{(2)}+O(\varepsilon), \label{RegTT}
\end{align}
where a scale breaking log appears. Notice that we could include a non conformal spin-1 sector, by the addition of the $\mathcal{N}$ term in \eqref{N}, that reveals a similar pattern as for $d=4$. Notice that this contribution vanishes linearly for $d\to 2$ since $B_0(p^2)$ is finite at $d=2$. In other words, the presence of corrections that are not Weyl invariant at tree level, once they are included as virtual corrections, do not invalidate the topological character of the trace anomaly.\\
We choose a {\em local} counterterm of the form 
\begin{align}
	S_{ct}=-\frac{1}{\varepsilon}\,\beta_c\,\int d^dx\, \sqrt{-g}\,\mu^{d-2}\,R
\end{align}
which is not Wey invariant for $d\neq 2$. 
The counterterm for the $TT$ in DR, expanded in $\epsilon=d-2$, is then given by 

\beqa
\braket{T^{\mu_1\nu_1}(p)T^{\mu_2\nu_2}(-p)}_{Count} =&& \nonumber \\
	-\frac{\beta_c\,p^2}{2\varepsilon}\,\left(\Pi^{\mu_1\nu_1\mu_2\nu_2}_{(2+2\varepsilon)}(p)-\frac{2\varepsilon}{(1+2\varepsilon)}\pi^{\mu_1\nu_1}(p)\pi^{\mu_2\nu_2}(p)\right)\times \nonumber \\
\times 	\left(1+\varepsilon \log\mu^2\right).&&\nonumber \\
\label{CountTT}
\eeqa
and the renormalized expression of the $TT$ is given by 
\begin{align}
	\braket{T^{\mu_1\nu_1}(p)T^{\mu_2\nu_2}(-p)}_{(d=2)}^{Ren}=&\frac{c(2)}{2}\Pi^{\mu_1\nu_1\mu_2\nu_2}_{(2)}\,p^2\log\left( \frac{p^2}{\mu^2}\right)\notag \\
	+p^2c^\prime(2)\,\Pi^{\mu_1\nu_1\mu_2\nu_2}_{(2)}+& c(2)\,p^2\pi^{\mu_1\nu_1}(p)\pi^{\mu_2\nu_2}(p).\label{TTRen}
\end{align}
where $c'(2)\equiv c'(d)\vert_{d=2}$. \\
It is evident, from the expression above, using \eqref{zerop}, that we are not going to have a scale anomaly and that the final expression of the correlator is given only by 
\beq
\label{fin}
\braket{T^{\mu_1\nu_1}(p)T^{\mu_2\nu_2}(-p)}_{(d=2)}^{Ren}=c(2)\,p^2\pi^{\mu_1\nu_1}(p)\pi^{\mu_2\nu_2}(p).
\eeq
in perfect agreement with the previous computation and the same anomaly. This can be verified immediately 
since  
\begin{equation}
	\pi^{\mu\nu}_{(d=2)}(p)\equiv\delta^{\mu\nu}-\frac{p^{\mu}p^{\nu}}{p^2}=\frac{n^\mu n^\nu}{p^2}
\end{equation}
and $n^2=p^2$, giving $\pi^\mu_\mu=1$. Therefore,  by tracing ${\mu_1\nu_1}$ in \eqref{fin}, we recover \eqref{ann}.
This shows that the choice of just a local counterterm is sufficient to generate the correct type of anomaly. 

\section{Conclusions}
The goal of our observations has been to point out that scale-breaking contributions are directly induced in the anomaly effective action by ordinary renormalization procedures. Such terms can be generated in an expansion around $d=4$ by subtractions which are directly borrowed from flat space in DR. They induce terms that are naturally Weyl invariant. A dilatation anomaly is present only for non topological counterterms.\\
 Our analysis illustrates also why a WZ action does not necessarily account for the dilatation anomaly. If we frame the WZ action in the context of a renormalization procedure, it is perfectly consistent to require that such action reproduces the trace anomaly, with the dilatation anomaly attributed to Weyl invariant terms which are not part of this (anomaly induced) action. The classification of anomaly as of type "A" or "B", the first preserving  scale invariance, and the second breaking this symmetry, is hidden by the approach. \\
At this point it is natural to ask whether it is acceptable that an anomaly induced action may miss the previous classification, since it does not account for the dilatation anomaly.  Our answer, based on the current analysis, is affirmative. 
An anomaly induced action is identified by solving a variational equation and it is not intended to reproduce 
the trace anomaly. It is built around the WZ consistency condition and misses the Weyl invariant terms which are part of the effective action. We have seen that there is something truly special in the way the WZ action is made finite, by relating two metrics , $g$ and $\bar{g}$. Log contribution are naturally hiding both in DR and in cutoff regularizations as soon as me perform a slight modifications of such regularizations, by finite subtractions. \\
From this perspective, the variational solution of the trace anomaly constraint (i.e. the anomaly actions), obtained 
by using different conformal decompositions, and their failure to reproduce the perturbative results, as discussed in \cite{Coriano:2022jkn}, force us to consider other and more urgent aspects and limits of these functionals, such as the inclusion of Weyl invariant terms. These are missing in such solutions and appear to be necessary in order 
to respect, in the correlation functions, the necessary conservation WIs. \\
This issue could be related to the difficulty of performing the flat spacetime limits of such actions starting from a curved metric background.\\ 
Finally, we have shown, in the case of the $TT$ 2-point function, that the topological and non topological components of the conformal anomaly, and their relation to the dilatation anomaly are also consistently taken care of by ordinary counterterms in DR.   
\vspace{0.4cm}

\centerline{\bf Acknowledgements} 
The work of C. C. and M.C. is funded by the European Union, Next Generation EU, PNRR project "National Centre for HPC, Big Data and Quantum Computing", project code CN00000013 and by INFN iniziativa specifica QFT-HEP.
  M. M. M. is supported by the European Research Council (ERC) under the European Union as Horizon 2020 research and innovation program (grant agreement No818066) and by Deutsche Forschungsgemeinschaft (DFG, German Research Foundation) under Germany's Excellence Strategy EXC-2181/1 - 390900948 (the Heidelberg STRUCTURES Cluster of Excellence). We thank Manuel Asorey, Emil Mottola, Ilya Shapiro, Luigi Delle Rose, Stefano Lionetti and Riccardo Tommasi for discussions.   


\providecommand{\href}[2]{#2}\begingroup\raggedright\endgroup

\end{document}